\documentclass[]{elsarticle}
\makeatletter
\def\ps@pprintTitle{%
	\let\@oddhead\@empty
	\let\@evenhead\@empty
	\def\@oddfoot{}%
	\let\@evenfoot\@oddfoot}
\makeatother
\usepackage{fancyhdr}

\usepackage{epsfig}
\usepackage{amsfonts}
\usepackage{amssymb, graphics, amsmath}
\usepackage{dsfont}
\usepackage{hyperref}
\usepackage{color}
\usepackage{pdflscape}
\usepackage{pifont}
\usepackage{graphicx} 
\usepackage{bm}
\usepackage{slashed}
\usepackage[compat=1.1.0]{tikz-feynman}
\usepackage{tikz}
\usepackage[compat=1.1.0]{tikz-feynhand}

\allowdisplaybreaks

\def\slashchar#1{\setbox0=\hbox{$#1$}           
	\dimen0=\wd0                                    
	\setbox1=\hbox{/} \dimen1=\wd1                  
	\ifdim\dimen0>\dimen1                           
	\rlap{\hbox to \dimen0{\hfil/\hfil}}            
	#1                                             
	\else                                          
	\rlap{\hbox to \dimen1{\hfil$#1$\hfil}}        
	/                                           
	\fi}      

\newcommand*{\rom}[1]{\expandafter\romannumeral #1}
\newcommand*{\Rom}[1]{\uppercase\expandafter{\romannumeral #1\relax}} 
\newcommand{\degg}{^{\circ}}

\begin{document}
\title{Differential Cross Section Predictions for PRad-\Rom{2} from
       Dispersion Theory}

\author[bonn]{Yong-Hui Lin}
\author[darmstadt,emmi]{Hans-Werner Hammer}
\author[bonn,fzj,tbilisi]{Ulf-G. Mei\ss{}ner}

\address[bonn]{Helmholtz-Institut f\"ur Strahlen- und
	Kernphysik and Bethe Center for Theoretical Physics,\\
	Universit\"at Bonn, D-53115 Bonn, Germany}
\address[darmstadt]{Technische Universit\"at Darmstadt, Department of Physics,
                    64289 Darmstadt, Germany}
\address[emmi]{ExtreMe Matter Institute EMMI, GSI Helmholtzzentrum f\"ur Schwerionenforschung GmbH,\\
               64291 Darmstadt, Germany}
\address[fzj]{Institute for Advanced Simulation, Institut f\"ur Kernphysik and
	J\"ulich Center for Hadron Physics,\\ Forschungszentrum J\"ulich,
	D-52425 J\"ulich, Germany}
\address[tbilisi]{Tbilisi State University, 0186 Tbilisi, Georgia}

\begin{abstract}
  We predict the differential cross sections for $e^-p$ and $e^+p$ elastic
  scattering in the PRad-\Rom{2} energy region. The prediction is based
  on form factors obtained in our previous high-precision analysis of space-
  and time-like data from dispersion theory and different sets of two-photon
  exchange corrections. We investigate the sensitivity of the cross sections
  to two-photon exchange effects and find
  that the differences between model calculations and phenomenological
  extractions of the two-photon corrections
  can not be resolved if the uncertainty in the form factors
  is taken into account.
\end{abstract}

\maketitle

\section{Introduction}

The nucleon form factors (NFFs)
are defined as the coefficients of the two independent
Dirac structures in the general form of the electromagnetic
$\gamma N N$ vertex function. The electric and magnetic Sachs
form factors $G_E$ and $G_M$ are scalar functions of the
four-momentum transfer $Q^2$ of the photon and encapsulate the
structure of the nucleon as seen by an electromagnetic probe. The
first experimental measurement of NFFs was carried out by Hofstadter and
collaborators at the Stanford High Energy Physics Laboratory
in 1953~\cite{Hofstadter:1953zz,Hofstadter:1956qs}. The understanding
of electromagnetic nucleon structure was further refined with the advent of
continuous beam electron facilities such as MAMI and
Jefferson Lab in the 1980's.
At low momentum transfers this line of investigations appeared to be complete
at the end of the last century. However,
a renaissance of the NFF studies in the 2000's was triggered by the
emergence of the so-called proton radius puzzle, based on the precise measurement of the
Lamb shift in muonic hydrogen, which led to the determination
of the proton charge radius at an unprecedented precision,
$r^P_E=0.8484(67)$~fm~\cite{Pohl:2010zza,Antognini:2013txn}.
This value differed by
$5\sigma$ from the CODATA value at that time~\cite{CODATA:2008}, which
was based on electron scattering experiments and measurements of the
Lamb shift in ordinary hydrogen.
For recent reviews of the NFFs see, e.g.,
Refs.~\cite{Denig:2012by,Pacetti:2014jai,Punjabi:2015bba}.

The charge radius of the proton is defined
as the derivative of its electric form factor at zero momenta transfer,
that is, $\langle (r_p^E)^2 \rangle= -6
dG_E^p/dQ^2(Q^2=0)$~\cite{Miller:2018ybm}.
This makes the first-principle calculation from Quantum Chromodynamics (QCD)
very difficult due to the non-perturbative nature of QCD in the low-energy region.
In lattice QCD, disconnected contributions to the isoscalar form factors
have proven to be a persistent challenge.
See, e.g., Ref.~\cite{Djukanovic:2021cgp} for the state of the art of
lattice QCD calculations.
A dispersion theoretical approach was proposed to analyze the electromagnetic
structure of the nucleon at the very beginning of experimental
investigations~\cite{Chew:1958zjr,Federbush:1958zz,Hohler:1976ax},
which includes all constraints from unitarity, analyticity and crossing
symmetry. It was further developed
in subsequent works~\cite{Mergell:1995bf,Belushkin:2006qa}
to include new experimental data and to
be consistent with the strictures from perturbative QCD at
very large momentum transfer \cite{Lepage:1980fj}. An important
recent development is the improved determination of the isovector spectral
functions from the two-pion continuum \cite{Hoferichter:2016duk},
based on an analysis of the Roy-Steiner equations for pion-nucleon
scattering \cite{Hoferichter:2015hva}.
A comprehensive review of the state of the art of dispersive analyses
of the nucleon electromagnetic form factors is given
in Ref.~\cite{Lin:2021umz}. The proton radius obtained in such
analyses agrees well with the ``small''  proton charge radius 
from muonic hydrogen and is quite robust.
It has remained around the value $\langle (r_p^E)^2 \rangle^{1/2}\simeq 0.84$~fm
since H\"ohler and collaborators pioneering work in the 1970's \cite{Hohler:1976ax}
despite many new experimental measurements and thus predates the
muonic hydrogen experiments,
see e.g. Ref.~\cite{Hammer:2019uab} for a discussion of the history.

Very recently, a comprehensive analysis of all latest $e^- p$ elastic scattering
(space-like, $Q^2>0$) and $e^+e^-\to N \bar{N}$ annihilation
(time-like, $Q^2<0$) data was performed
in Ref.~\cite{Lin:2021xrc} within the dispersion theoretical framework. In
the space-like region, the momentum transfer range
$Q^2=0.000215$-$0.977\,\rm (GeV/c)^2$
is covered, which consists mainly of the cross section data from
MAMI-C~\cite{A1:2013fsc} and PRad~\cite{Xiong:2019umf}.
In the time-like region it includes data in the
region $|Q^2|=3.52$-$20.25\,\rm (GeV/c)^2$ that is mainly composed of the
effective form factor data from
BES\Rom{3}~\cite{BESIII:2021rqk,BESIII:2021dfy} and
BaBar~\cite{BaBar:2013ves}\footnote{Only the data basis covering the main
energy range are listed here. The full data basis used in that
comprehensive analysis can be found in Ref.~\cite{Lin:2021xrc}.}.
A set of nucleon form factors that describes all existing experimental data both
in the space- and time-like regions over the full range of momentum transfers
was obtained. In addition, the extracted form factors are also
consistent with the
measurements of Lamb shift and hyperfine splittings in muonic
hydrogen as discussed in~\cite{Lin:2021xrc}. Starting from this
set of NFFs, one can predict the electromagnetic properties of
nuclei and experimental cross sections that will be 
be measured in the future.

Recently, two upgraded projects based on the PRad setup for the $e^- p$ and
$e^+ p$ elastic scattering were proposed
in Refs.~\cite{PRad:2020oor} and \cite{Hague:2021xcc}, respectively.
The former is also called PRad-\Rom{2}. In this work, we
calculate the differential cross sections for the $e^-p $ and $e^+p$ elastic
scattering in the energy region where the PRad-\Rom{2}
experiment will run
based on the high-precision nucleon form factors obtained in
Ref.~\cite{Lin:2021xrc}. The results 
serve as predictions for the upcoming PRad-\Rom{2} and $e^+ p$ scattering
measurements.

\section{Formalism}\label{sec:form}

Here, we briefly summarize the underlying formalism, which is detailed in
Ref.~\cite{Belushkin:2006qa,Lorenz:2014yda,Lin:2021umz}. When working with the
one-photon-exchange assumption, the differential cross section for $e^-p$
scattering can be expressed through the Sachs form factors
$G_E$ and $G_M $ as
\begin{equation}\label{eq:xs_ros}
\frac{d\sigma_{1\gamma}}{d\Omega} = \left( \frac{d\sigma}
{d\Omega}\right)_{\rm Mott} \frac{\tau}{\epsilon (1+\tau)}
\left[G_{M}^{2}(Q^2) + \frac{\epsilon}{\tau} G_{E}^{2}(Q^2)\right]\, ,
\end{equation}
where $\epsilon = [1+2(1+\tau)\tan^{2} (\theta/2)]^{-1}$ is the virtual photon
polarization,
$\theta$ is the electron scattering angle in the laboratory frame,
$\tau = Q^2/4m_N^2$, with $Q^2>0$
the four-momentum transfer squared and $m_N$ the nucleon mass. It is
well known as the Rosenbluth formula. Moreover,
$({d\sigma}/{d\Omega})_{\rm Mott}$ is the Mott cross section, which describes
scattering off a point-like spin-1/2 particle.
To be compatible with the experimental data,
the theoretical calculation must go beyond the one-photon approximation and in
principle should consider all higher order corrections which are not negligible
at the accuracy of experiments. However, these corrections, mainly composed of
the radiative corrections, have usually been subtracted when the
experiments present their measurements. The majority of them are estimated with
the model-independent formalism which been widely investigated in the
literature~\cite{deCalan:1990eb,Maximon:2000hm,Weissbach:2008nx,Akushevich:2015toa,Arbuzov:2015vba,Bucoveanu:2018soy}.
In contrast, the two-photon-exchange (TPE) contribution~\cite{Afanasev:2017gsk},
which contains a relatively large uncertainty due to its model-dependent nature, 
is treated differently in various experiments. For example, MAMI-C uses a combination
of the soft-photon limit~\cite{Mo:1968cg,Maximon:2000hm} and a two-parameter
empirical parametrization to account for the TPE correction, while a
dispersive formalism~\cite{Tomalak:2014sva,Tomalak:2015aoa} is used in the PRad experiment.
In the present work, the differential cross sections of $e^\pm p$ scattering are
calculated in the one-photon approximation Eq.~\eqref{eq:xs_ros}
and TPE corrections are applied. The sensitivity of the cross section
to different sets of TPE corrections is investigated.   

The leading TPE contribution is given by the interference between the one-photon
and two-photon exchange amplitudes, $\mathcal{M}_{1\gamma}$ and $\mathcal{M}_{2\gamma}$, respectively.
It can be accounted for by including a modifying factor in Eq.~\eqref{eq:xs_ros}.
In the case of $e^- p$ scattering, the cross section including two-photon corrections can
be written as
\begin{equation}
  \label{eq:2pc}
  \left(\frac{d\sigma_{\rm corr}}{d\Omega}\right)_{e^- p} = \frac{d\sigma_{1\gamma}}{d\Omega}(1 + \delta_{2\gamma})~,\quad 
  ~\delta_{2\gamma}
=\frac{2\text{Re}(\mathcal{M}^{*}_{1\gamma}\mathcal{M}_{2\gamma}^{})}{|\mathcal{M}_{1\gamma}|^2}\,(1+\mathcal{O}(\alpha))~.
\end{equation}
For the two-photon exchange amplitude $\mathcal{M}_{2\gamma}$, we adopt the calculations of
Refs.~\cite{Lorenz:2014yda,Lin:2021umz}. 
The box and crossed box diagrams are evaluated in the framework of a hadronic model with both the nucleon and
$\Delta$-resonance contributions included~\cite{Lorenz:2014yda}. The update made in the present work
refers to the one-photon amplitude $\mathcal{M}_{1\gamma}$ where the latest dispersive NFFs from
Ref.~\cite{Lin:2021xrc} are used.
This set of TPE corrections is representative for such model
calculations and other calculations give consistent
results~\cite{Afanasev:2017gsk}.

The energy parameters of PRad-\Rom{2} are taken from the proposal~\cite{PRad:2020oor}:
three incident electron beams with energy $E_e=0.7$, $1.4$, $2.1$~GeV at very small scattering angles
$\theta_e$ from $0.5^{\degg}$ to $7^{\degg}$, corresponding to $Q^2=4\times10^{-5}$ to $6\times10^{-2}~\rm{(GeV/c)^2}$.
This new experiment is designed to achieve a factor of 3.8 reduction in the overall experimental uncertainties
of the extracted proton radius compared to PRad. With such an improvement of precision, it is expected to address
possible systematic difference between the radius from $e^-p$ and $\mu$H measurements. The proposed
$e^+p$ scattering would use the same setup as the PRad-\Rom{2} experiment~\cite{Hague:2021xcc}.
The cross section of $e^+p$ scattering is exactly same as that of $e^-p$, except that the factor
$(1 + \delta_{2\gamma})$ in Eq.~(\ref{eq:2pc}) must be changed to $(1 - \delta_{2\gamma})$.

\section{Results}
\begin{figure}[htbp]
\begin{center}
\includegraphics[width=0.7\textwidth]{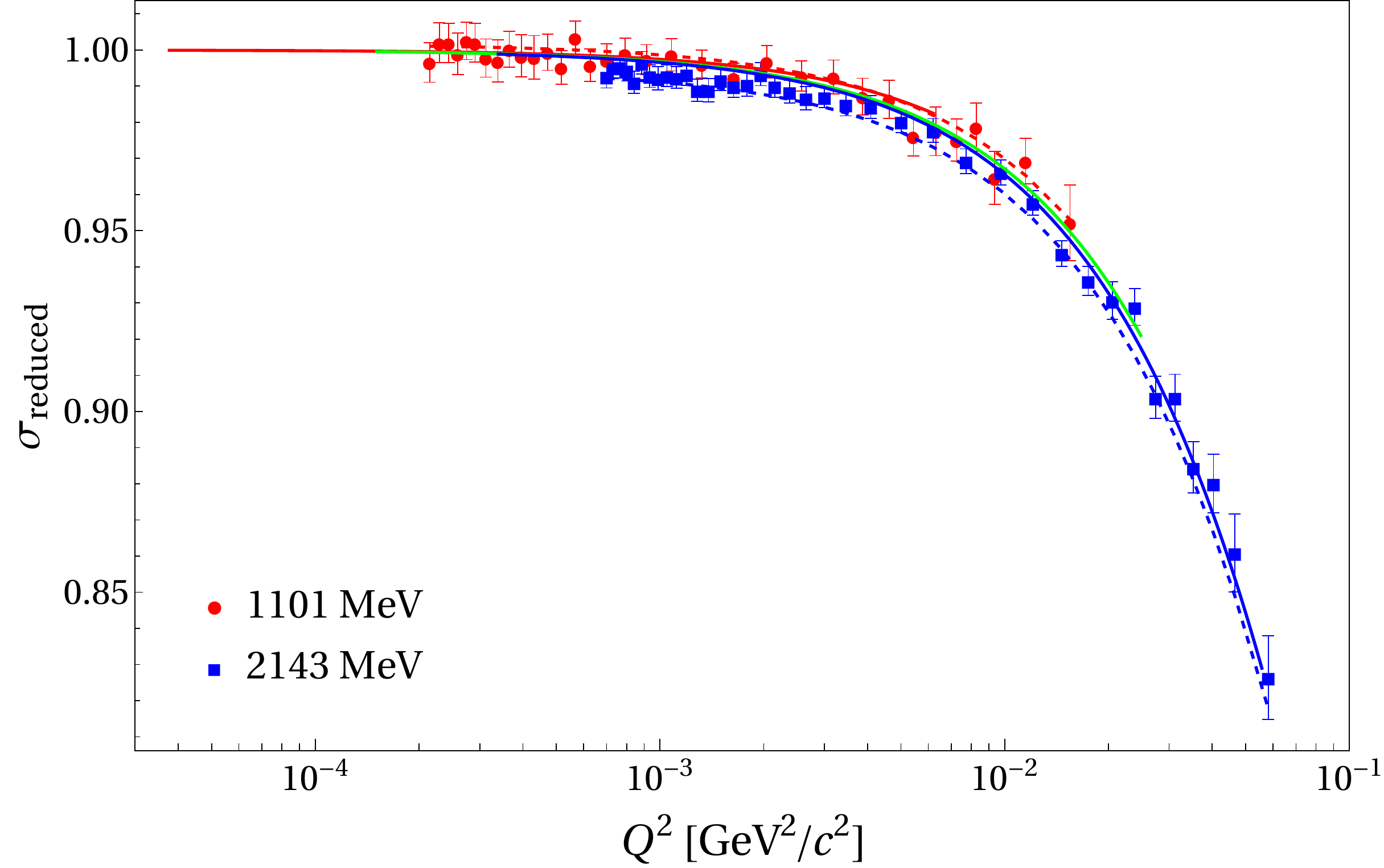}
\caption{Reduced cross section $\sigma_{\rm reduced}$ in the energies
  regions probed by PRad and PRad-\Rom{2}. The PRad data~\cite{Xiong:2019umf}
  is shown for comparison by the red solid circles ($1.1$~GeV data) and
  blue solid squares ($2.2$~GeV data). The red and blue dashed lines show
  our corresponding results for PRad. The red, green and blue solid lines
  are our predictions for the $0.7$, $1.4$ and $2.1$~GeV energy beams
  of PRad-\Rom{2}, respectively.
\label{fig:PRad1vs2}}
	\end{center}
\end{figure}

Now let us move to the differential cross sections.
For easier comparison with PRad~\cite{Xiong:2019umf} and the
PRad-\Rom{2} proposal~\cite{PRad:2020oor},
we calculate the reduced cross section $\sigma_{\rm reduced}$ defined as
\begin{align}
  \sigma_{\rm reduced}&\equiv \left(\frac{d\sigma}{d\Omega}\right)_{ep}
  \times \left[
          \left(\frac{d\sigma}{d\Omega}\right)_{\rm Mott}
          \frac{E'/E}{1+\tau}\right]^{-1}\nonumber\\
	&= \frac{E}{\epsilon E'}
	\left[\tau G_{M}^{2}(Q^{2}) + {\epsilon} G_{E}^{2}(Q^{2})\right]~,
\end{align}
where $E$ and $E'$ are the incoming and outgoing electron energies in the
laboratory frame.
In Fig.~\ref{fig:PRad1vs2}, we show $ \sigma_{\rm reduced}$ in the
energy regions probed by PRad and PRad-\Rom{2}. Although there is a
large overlap in the energy region measured in the  PRad-\Rom{2} and
PRad experiments evident in Fig.~\ref{fig:PRad1vs2}, the uncertainty reduction
and lower $Q^2$ extension proposed in PRad-\Rom{2} are extremely important
for the extraction of the proton radius. 

\begin{figure}[tbp]
\begin{center}
\includegraphics[width=0.7\textwidth]{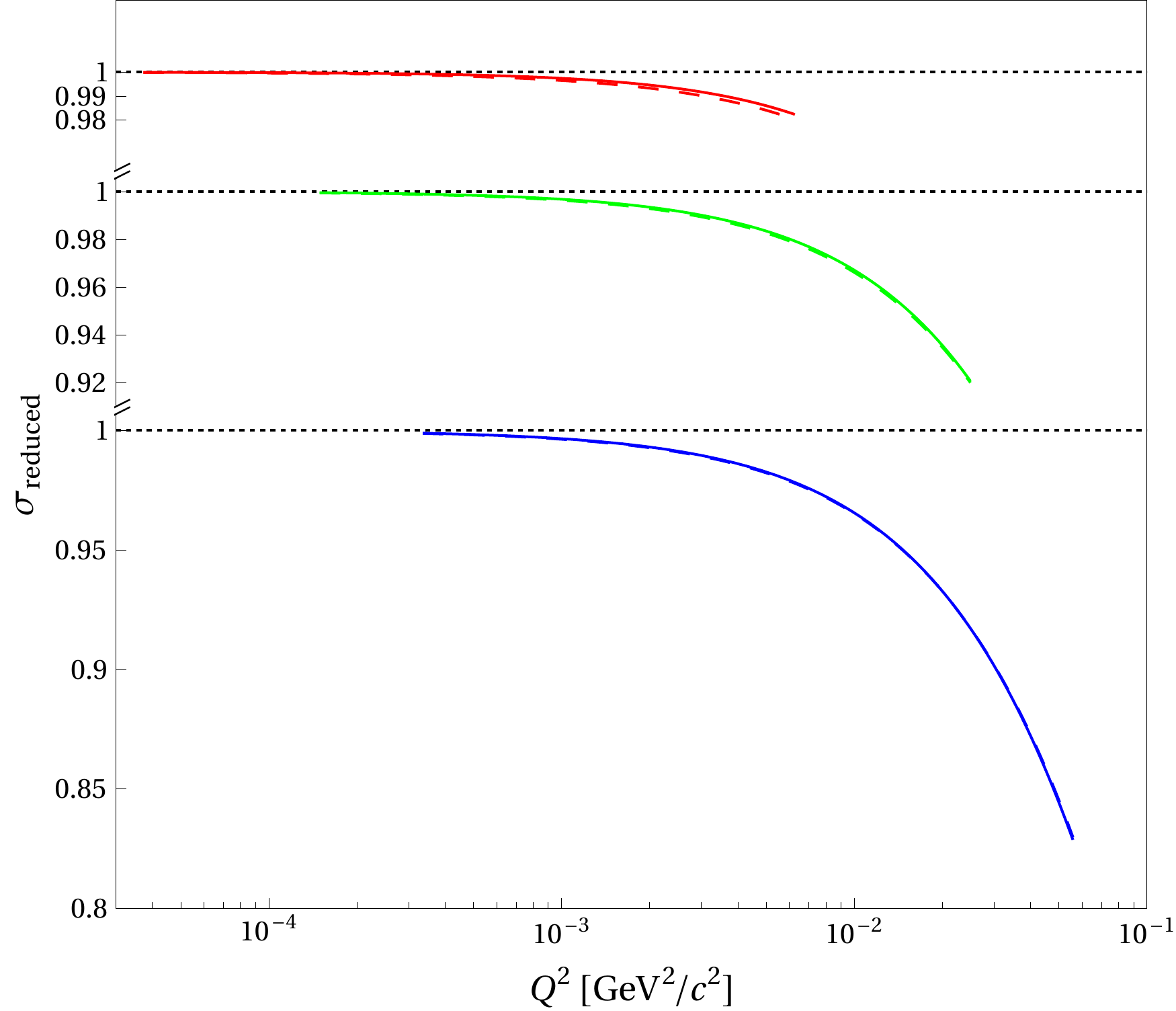}
\caption{Reduced cross section $\sigma_{\rm reduced}$ for the
  PRad-\Rom{2} (solid) and positron-proton scattering (dashed) experiments.
  The red, green and blue color represent the results for the $0.7$, $1.4$
  and $2.1$~GeV energy beams.
			\label{fig:PRad2epvspp}
		}
	\end{center}
\end{figure}

\begin{figure}[htbp]
\begin{center}
\includegraphics[width=0.45\textwidth]{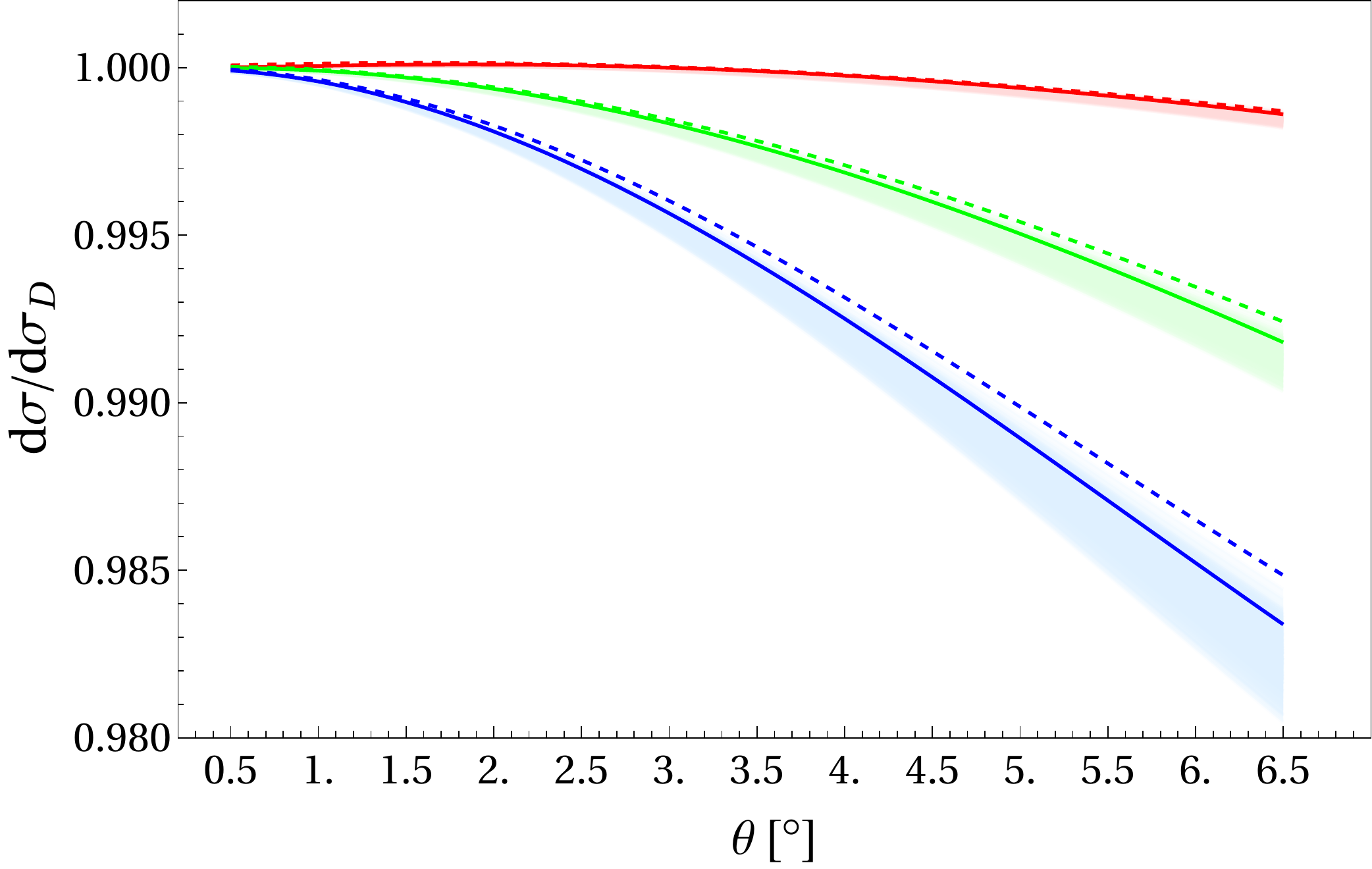}
\includegraphics[width=0.45\textwidth]{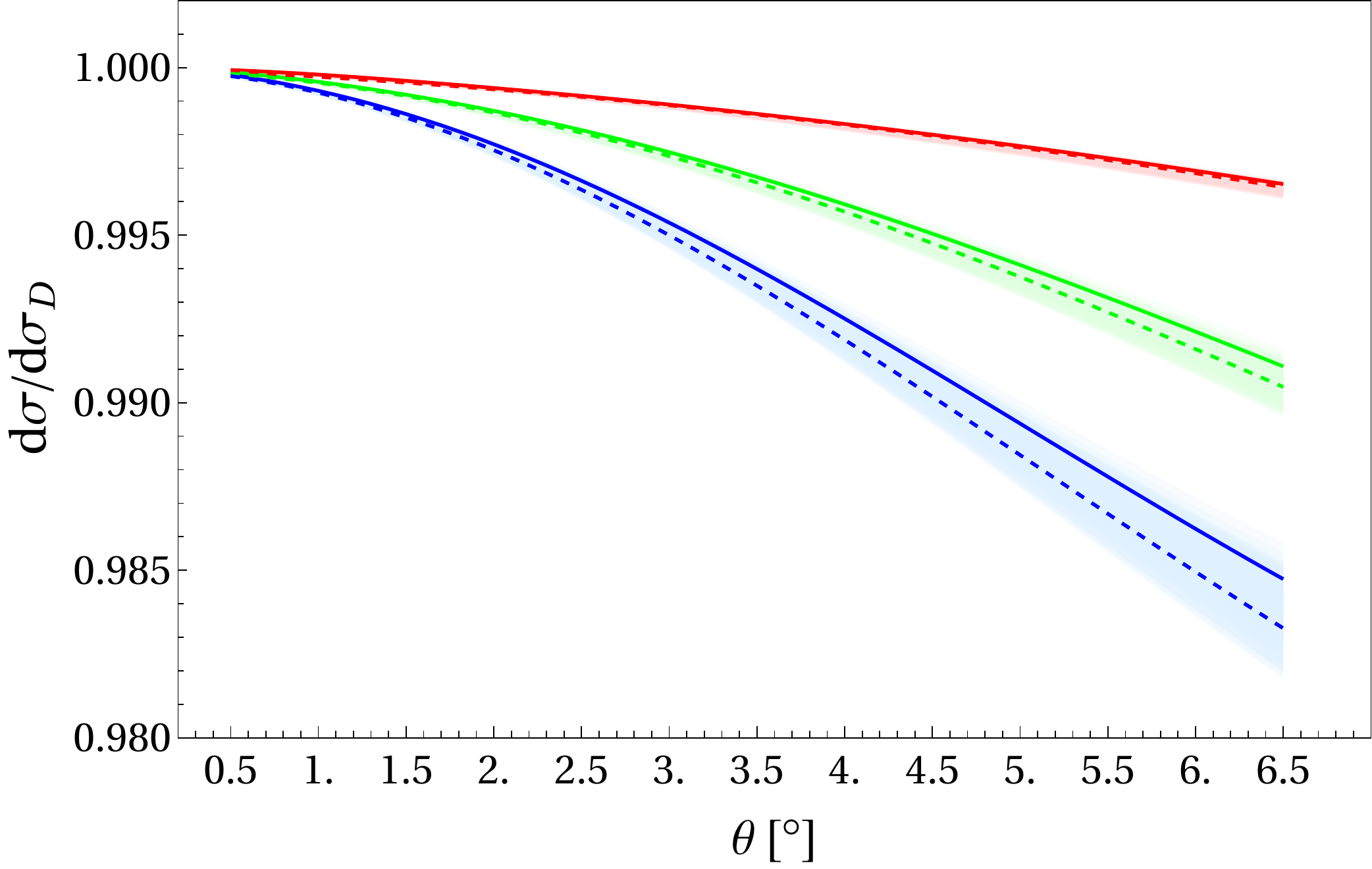}
\caption{Differential cross sections normalized to the cross section for
  dipole Sachs FFs for the PRad-\Rom{2} (left panel) and $e^+ p$ scattering
  (right panel) experiments. The red, green, and blue solid lines are the
  results with the box-graph TPE for the $0.7$, $1.4$ and $2.1$~GeV energy
  beams, respectively. The corresponding dashed lines show the results with
  the phenomenological TPE from Bernauer et al.~\cite{A1:2013fsc}. The
  uncertainties from the NFFs are presented as the shaded bands.
			\label{fig:dcsPRad2}
		}
	\end{center}
\end{figure}

The difference in the reduced cross sections for $e^+p$ and $e^-p$
scattering for the  $0.7$, $1.4$, and $2.1$~GeV energy beams is shown
in Fig.~\ref{fig:PRad2epvspp}. It is only due to TPE effects and
thus rather small.
Furthermore, we investigate the effect of TPE corrections on the
predicted differential cross sections in detail. Two different schemes for
TPE corrections, the box-graph model from
Refs.~\cite{Lorenz:2014yda,Lin:2021umz}, which is representative
for hadronic model calculations of TPE,
and the phenomenological parametrization from Bernauer et al.~\cite{A1:2013fsc}
are compared in Fig.~\ref{fig:dcsPRad2} as a function of the
scattering angle $\theta$. The cross sections are
normalized to ones calculated using the dipole Sachs FFs to make the
deviation among various TPE corrections transparent.
Within the uncertainties of the NFFs, the cross sections with the box-graph
model TPE are in agreement with those calculated by the parametrization
obtained in Ref.~\cite{A1:2013fsc}\footnote{In principle, the uncertainties in the NFFs should also enter the box-graph TPE calculation as discussed
in Sec.~\ref{sec:form}. However, the magnitude of the uncertainty in the TPE
correction is a second order effect and thus quite small.}.
The largest deviations occur for larger angles corresponding
to the higher $Q^2$-range as expected.
It turns out that the uncertainty emerging from the model-dependent TPE
corrections becomes a secondary effect when it is compared with the NFF
uncertainties in the PRad-\Rom{2} energy region. The high-precision
data from PRad-\Rom{2} thus will provide strong constraints on the nucleon form
factors and help us improve the precision of our dispersive NFFs
in Ref.~\cite{Lin:2021xrc}.

Another interesting issue that the positron-proton scattering experiment
would address is the ratio of cross sections for $e^+ p$ and $e^- p$
scattering. It will give direct access to the TPE corrections. This can
be seen clearly from Fig.~\ref{fig:dcsPRad2}, where the sign of the
deviation between two different TPE calculations is opposite for
the $e^- p$ and $e^+ p$ scattering processes. This implies a large
effect in the ratio of cross sections. We will come back to this
issue in a future study (see also Ref.~\cite{Afanasev:2017gsk} for
a recent review).

\section{Summary}

In this paper, we present a prediction for the differential cross sections
for the upcoming PRad-\Rom{2} and positron-proton scattering experiments
based on the dispersive nucleon form factors obtained in
Ref.~\cite{Lin:2021xrc}. Moreover, the effect of two-photon-exchange
corrections is also discussed. We find that the uncertainty caused by
the model-dependent nature of TPE corrections is a secondary contribution
when compared to the uncertainties in the NFFs errors for the
PRad-\Rom{2} setup. The high-precision PRad-\Rom{2} data will provide
strong constraints on the nucleon form factors and improve our
knowledge of the electromagnetic structure of the nucleon. Finally,
the theoretical uncertainties of the two-photon-exchange corrections
need to be investigated further together with the positron-proton
scattering data that will be available in the near future.

\section*{Acknowledgements}

This work of UGM and YHL is supported in
part by  the DFG (Project number 196253076 - TRR 110)
and the NSFC (Grant No. 11621131001) through the funds provided
to the Sino-German CRC 110 ``Symmetries and the Emergence of
Structure in QCD",  by the Chinese Academy of Sciences (CAS) through a President's
International Fellowship Initiative (PIFI) (Grant No. 2018DM0034), by the VolkswagenStiftung
(Grant No. 93562), and by the EU Horizon 2020 research and innovation programme, STRONG-2020 project
under grant agreement No 824093. 
HWH was supported by the Deutsche Forschungsgemeinschaft (DFG, German
Research Foundation) -- Projektnummer 279384907 -- CRC 1245
and by the German Federal Ministry of Education and Research (BMBF) (Grant
no. 05P21RDFNB).  Further, this project has received funding from the European Research Council
(ERC) under the European Union's Horizon 2020 research and innovation programme
(grant agreement No. 101018170).



\bibliographystyle{apsrev4-1}
\bibliography{refs}

\end{document}